\documentclass{emulateapj}

\newcommand{\sub}[2]{\ensuremath{#1_{\mathrm{#2}}}}
\newcommand{\super}[2]{\ensuremath{#1^{\mathrm{#2}}}}
\newcommand{\unit}[2]{\ensuremath{\textrm{#1}^{#2}}}
\newcommand{\vect}[1]{\mathbf{#1}}
\newcommand{\ph}[1]{\phantom{#1}}

\newcommand{\shh}{S15}

\usepackage{graphicx} 
\usepackage{epstopdf} 
\usepackage{amsmath}
\usepackage{hyperref}
\usepackage{natbib}
\usepackage{color}
\definecolor{hcitecolor}{RGB}{0,127,0}
\hypersetup{%
  citecolor=hcitecolor,%
  linkcolor=hcitecolor,%
  urlcolor=hcitecolor%
}%

\shorttitle{\textsc{Inferring the Galactic potential with Gaia and friends}}
\shortauthors{\textsc{Sanderson 2015}}

\begin{document}
\title{Inferring the Galactic potential with Gaia and friends:\\ synergies with other surveys}
\author{Robyn E. Sanderson\altaffilmark{1}} 
\affil{Department of Astronomy, Columbia University, New York, NY 10027, USA}
\altaffiltext{1}{NSF Astronomy and Astrophysics Postdoctoral Fellow}
\email{robyn@astro.columbia.edu}
\slugcomment{version revised \today} 

\begin{abstract}
In the coming decade the Gaia satellite will precisely measure the positions and velocities of millions of stars in the Galactic halo, including stars in many tidal streams. These streams, the products of hierarchical accretion of satellite galaxies by the Milky Way (MW), can be used to infer the Galactic gravitational potential thanks to their initial compactness in phase space. Plans for observations to extend Gaia's radial velocity (RV) measurements to faint stars, and to determine precise distances to RR Lyrae (RRLe) in streams, would further extend the power of Gaia's kinematic catalog to characterize the MW's potential at large Galactocentric distances. In this work I explore the impact of these extra data on the ability to fit the potential using the method of action clustering, which statistically maximizes the information content (clumpiness) of the action space of tidal streams, eliminating the need to determine stream membership for individual stars. Using a mock halo in a toy spherical potential, updated post-launch error models for Gaia, and estimates for RV and distance errors for the tracers to be followed up, I show that combining either form of additional information with the Gaia catalog greatly reduces the bias in determining the scale radius and total mass of the Galaxy, compared to the use of Gaia data alone. 
\end{abstract}

\section{Introduction}

The strategy of determining the gravitational potential of the Galaxy by modelling its tidal streams is a popular one, whose practical applications date back to the first measurements of the proper motion (PM) of the Magellanic Clouds \citep{1994MNRAS.270..209M,1995ApJ...439..652L} and the discovery of the Sagittarius dwarf galaxy \citep{1994Natur.370..194I} and its associated tidal stream \citep{1995ApJ...451..598J,1996ApJ...458L..13M}. The Sloan Digital Sky Survey (SDSS) confirmed the existence of many more tidal streams in our galaxy \citep{2002ApJ...569..245N} and both SDSS and more recent surveys continue to find new structures as their footprint expands \citep{2014ApJ...787...19M,2014ApJ...791....9S,2014MNRAS.443L..84B}. However, attempts to constrain the potential with known streams have so far been information-limited due to the difficulty of measuring sufficient of the 6 components of the phase-space positions of stream stars and of determining the membership of particular stars in a stream. As a result most attempts have focused on modeling a single stream at a time, which makes it difficult to simultaneously constrain both the properties of the progenitor of the stream and the Galactic potential. 

The Gaia astrometric mission \citep{2001A&A...369..339P}, launched at the end of 2013 and now taking data, will address the relative dearth of phase-space information for stars in tidal streams by measuring precise parallaxes and PMs for 1 billion stars to $V\sim20$ \citep{2014EAS....67...23D} and radial velocities (RVs) for the stars with $V \lesssim 15$\footnote{\url{http://www.cosmos.esa.int/web/gaia/science-performance}}. This unprecedented catalog will thus provide full phase-space information for many stream stars, but without membership information. Furthermore the faintest stars, which will lack radial velocity information, are likely to be the most distant and therefore the most interesting from the standpoint of measuring the Galactic mass distribution. 

To maximize the power of the Gaia survey (as well as other planned space-based surveys such as Euclid), complementary spectroscopic surveys are being planned in both the Northern and Southern hemispheres that, among their other goals, will obtain both radial velocities and chemical abundances for some fraction of stars with Gaia astrometry. In the north two such surveys have been planned: the WEAVE survey\footnote{\url{http://www.ing.iac.es/weave/science.html}}, hosted on the William Herschel telescope at La Palma \citep{Dalton2012}, and the DESI survey\footnote{\url{http://desi.lbl.gov}} \citep{doi:10.1117/12.2057105} at the Mayall telescope on Kitt Peak. The southern counterpart is the 4MOST survey\footnote{\url{http://4most.eu}} on the VISTA telescope at ESO \citep{DeJong2012}. These instruments all have similar performance specifications and are intended to complement one another in the different hemispheres. 

\begin{figure}
\begin{center}
\plotone{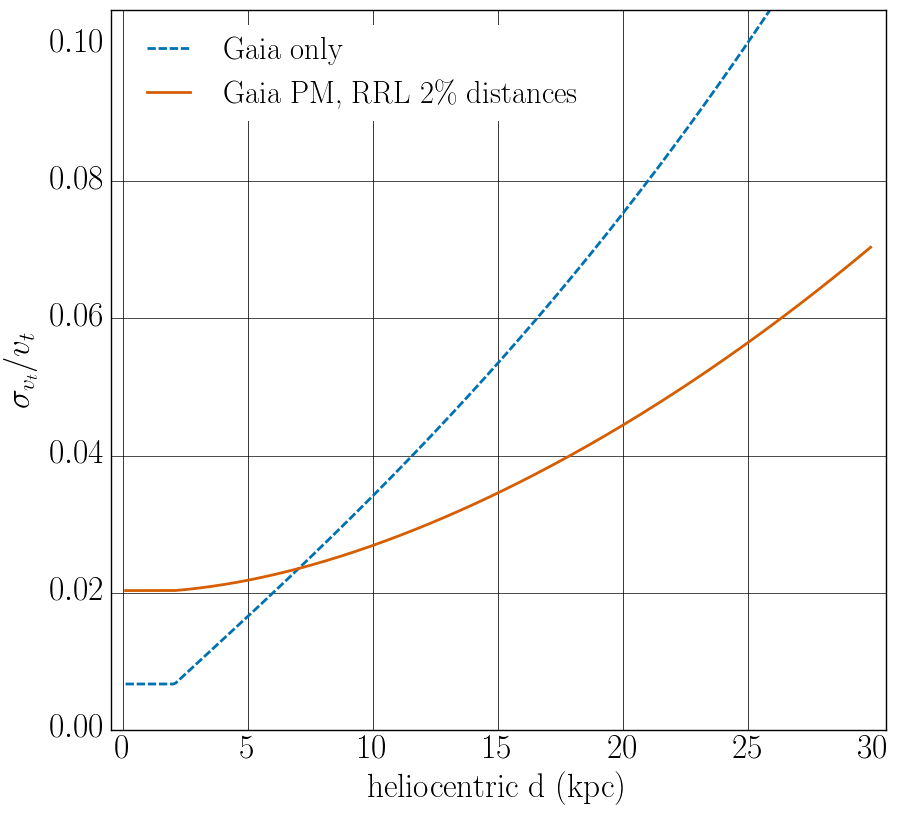}
\caption{Relative transverse velocity error as a function of distance for RRLe ($M_V=0.65$, $(V-I)=0.58$) using Gaia PMs and parallaxes (blue dashed line) and using Gaia PMs with 2\% distance measurements from the period-luminosity relation (red solid line).}
\label{fig:vterror}
\end{center}
\end{figure}

Another promising complement to the Gaia catalog is the capability to measure extremely accurate distance moduli for RR Lyrae (RRLe) by observing them in the infrared, where the scatter in the period-luminosity ($P$-$L$) relation is significantly smaller than at visible wavelengths \citep[][and references therein]{2012ApJ...744..132M}.  As pointed out by \citet{2013ApJ...778L..12P}, this scatter could be as low as 2\% in the 3.6-micron band observed by Spitzer, and the resulting super-accurate distances could generate powerful new constraints on the Galactic potential. Because Gaia measures angular velocity on the sky, accurate distances are the key to accurate tangential velocity measurements. Figure \ref{fig:vterror} illustrates the improvement in the tangential velocity ($v_t$) error when obtaining 2\% distances from the $P$-$L$ relation rather than using Gaia parallaxes, assuming typical values for the absolute magnitude ($M_V$) and $V-I$ color of RRLe. Whereas with Gaia alone the error on $v_t$ exceeds ten percent by about 25 kpc (heliocentric), with improved RRLe distances the relative error reaches ten percent at around 40 kpc, and increases more slowly with distance. Since RRLe are intrinsically bright stars, obtaining accurate distances nearly doubles the reach of the Gaia catalog in terms of transverse velocity accuracy. Observations of RRLe with Spitzer can reach to 40 kpc assuming reasonable integration times \citep{2013ApJ...778L..12P}. If the ground-based spectroscopic surveys discussed above obtain radial velocities for distant RRLe (a reasonable assumption, since they have a low sky density and would require at most a few fibers per field), the kinematic substructures traced by these stars could be viewed in exquisite detail in six dimensions, well into the outer halo.

The forthcoming avalanche of kinematic information for tidal stream stars calls for new approaches to constraining the Galactic potential that can use this information to full advantage. In \citet[][hereafter \shh]{Sanderson2014}, we outlined one such method that works by maximizing the information contained in the space of the stars' actions. The actions of stream stars are expected to be clustered since each stream starts as part of a gravitationally bound object that occupies a small phase-space volume, relative to that of the MW as a whole. Although it requires full six-dimensional phase-space information for all the stars used, this method does not require assigning stars to particular streams (or even that all the stars be members of a tidal stream), so even stars in previously undiscovered streams can contribute to the fit. The method also produces a robust measurement of the present-day potential for streams in a time-evolving, lumpy cosmological halo (Sanderson, Hartke, \& Helmi, in prep). The main points of our method are reviewed in Section \ref{sec:kld} of this paper. 

\shh\ tested the performance of the action-clustering method, using a mock halo in a toy potential, for stars whose full 6D phase-space positions would be observed by Gaia alone, and found that the distance range of these stars (out to about 20 kpc from the Galactic center) was too small to get accurate independent measurements of the total mass and scale radius of the toy potential, although the method accurately recovered the enclosed mass at the average stellar radius to good precision. The observational errors expected from Gaia, especially the parallax errors, serve to blur the action-space clumps and contribute to the bias on the total mass and scale as well as increasing the uncertainties on these parameters. These results, however, were based on the pre-launch performance estimates for the spacecraft; the post-launch performance estimates released this year indicate that RVs will not go as faint as originally planned, decreasing the distance range over which Gaia will obtain full phase-space coordinates. The astrometric errors are now also projected to be slightly worse for the faint stars whose influence is so crucial to understanding the gravitational potential at large radii.

This paper considers how two types of ground-based follow-up that would contribute to the completion of the 6D Gaia kinematic catalog will affect our ability to determine the total mass and scale of the Galaxy: radial velocities for distant KIII giants beyond the magnitude limit of the Gaia RV spectrophotometer, and accurate distances for RRLe. Each of these tracers is bright enough for Gaia to measure PMs at more than 50 kpc from the Galactic center, addressing the distance-range issue. The accurate distances available for RRLe address the primary way in which observational errors interfere with the fit (through the errors on the parallax). For consistency, the same mock halo is used as in \shh, with a resampling to account for the estimated number of RRLe rather than KIII giants in the tidal streams making up the halo, as described in Section \ref{sec:mockhalo} of this work. Section \ref{sec:mockhalo} also describes the estimated observational errors for these two tracers, which have been updated to the Gaia post-launch error model. Sections \ref{sec:results} and \ref{sec:discussion} present and discuss the results.

\section{Potential fitting with action-space clustering}
\label{sec:kld}
The potential-fitting method was developed in \shh, and simultaneously fits multiple streams to a common potential by maximizing the clustering of the stream stars in action space. The clustering is measured statistically using the Kullback-Leibler divergence (KLD) or relative entropy \citep{Kullback1951},
\begin{equation}
 \sub{D}{KL}(p : q) \equiv \int p(\vect{x}) \log \frac{p(\vect{x})}{q(\vect{x})} d\vect{x},\footnote{The logarithm can be any base; for this work we use the natural (base $e$) log, for which the units of the KLD are known as ``nats".  }
\end{equation}
which measures the difference in the information content of two distributions, here represented as $p$ and $q$. In this case the gravitational potential is represented in terms of some parameters $\vect{a}$, and the distributions to be compared are of the actions $\vect{J}$ of the stars in the fitting sample, which are a function of these potential parameters and the stars' phase-space positions $\vect{w}$. The distribution of actions obtained for a given set of potential parameters will be denoted $f_{\vect{a}}(\vect{J})$; all the $\vect{J}$ are calculated from the same underlying set of $\vect{w}$ and only the parameters $\vect{a}$ change. The KLD treats every star equally when measuring the clustering, and thus does not require assigning stream membership to the stars, or knowledge of how many streams are in the sample. There is also no need for every star in the sample to belong to a stream, though as discussed in \shh\ the distribution must not be dominated by a central smooth population.

The fit proceeds in two steps, each using the KLD in a different way. Step I identifies the best fit by calculating the KLD between the distribution of actions for a given set of parameters, $f_{\vect{a}}(\vect{J})$, and the distribution produced by shuffling the values of the actions relative to each other, $\super{f_{\vect{a}}}{shuf}(\vect{J})$:
\begin{equation}
\label{eq:dkl1}
 \sub{\super{D}{I}}{KL} =\int f_{\vect{a}}(\vect{J}) \log \frac{f_{\vect{a}}(\vect{J})}{\super{f_{\vect{a}}}{shuf}(\vect{J})} d^3\vect{J}.
\end{equation}
Comparing the distribution with its shuffled version is equivalent to comparing it with the product of the marginal distributions along each action dimension, and automatically accounts for the proper range of actions. This type of KLD between a distribution and the product of its marginals is known as the mutual information. The best-fit parameters, $\sub{\vect{a}}{0}$, are the ones with the largest value of $\sub{\super{D}{I}}{KL}$. 

To estimate the uncertainty on the best-fit parameters (Step II) the KLD is used again, to compare the action distribution obtained for the best-fit parameters $\sub{\vect{a}}{0}$ to the distribution obtained for another $\sub{\vect{a}}{trial}$:
\begin{equation}
\label{eq:dkl2}
\sub{\super{D}{II}}{KL} = \int f_{\sub{\vect{a}}{0}}\left(\vect{J}\right)  \log \frac{f_{\sub{\vect{a}}{0}}\left(\vect{J}\right)}{f_{\sub{\vect{a}}{trial}}\left(\vect{J}\right)}\ d^3\vect{J}
\end{equation}
This version of the KLD is related to the probability that the action distribution obtained using trial parameters $\sub{\vect{a}}{trial}$ represents the distribution obtained using the best-fit parameters $\sub{\vect{a}}{0}$, averaged over the stars in the sample. In \shh\ we showed that $\sub{\super{D}{II}}{KL}$ can be used to give uncertainties via the relation:
\begin{equation}
\label{eq:kld2interp}
\sub{\super{D}{II}}{KL} = \Big\langle \log \frac{ \mathcal{P}( \vect{a}_0 | \vect{J} ) }{ \mathcal{P}(\sub{\vect{a}}{trial}|\vect{J}) }\Big\rangle_{\vect{J}}.
\end{equation}
That is, $\sub{\super{D}{II}}{KL}$ is the log conditional probability of $\sub{\vect{a}}{trial}$ relative to $\vect{a}_0$, averaged over the action distribution given by Step I. Thus if a given $\sub{\vect{a}}{trial}$ has  $\sub{\super{D}{II}}{KL} = -\log\left(1/2\right)$, then we say that set of parameters is half as likely as the best fit. For reference, if the distribution around the best fit point could be represented by a Gaussian then the contour where the probability is $e^{-1/2} \approx$ 60\% of the maximum occurs $1\sigma$ away from the max; the $2\sigma$ and $3\sigma$ contours are where the relative probability is $e^{-2}$ and $e^{-9/2}$ respectively. Since we use natural logs to calculate the KLD, the analogous reference levels are $\sub{\super{D}{II}}{KL} = (-1/2, -2, -9/2)$. Although the assumption of Gaussian uncertainties is not necessarily applicable here, these levels assist in interpreting the range of acceptable parameters.

Both KLD values are computed using a grid-based modified Breiman kernel density estimator \citep{2011A&A...531A.114F}, which infers the density at a set of grid points spanning the 3D action space. The KLD in each case is calculated by performing a direct Riemann sum over the grid; the details of the numerical implementation are the same as in \shh. For the KIII giant samples a 256x256x2048 grid is used, with the finer sampling in the $J_r$ dimension (the only one that changes with the potential parameters). This resolution produces KLD values converged to less than 5 percent. For the RRLe sample, which contains far fewer stars, a $128^3$ grid is sufficient to converge the KLD to a few percent; increasing the grid resolution would severely oversample the density distribution.

Using the numerical KLD estimator for Step I, the two-dimensional parameter space of the potential is searched on a grid over the range $9<\log(M/M_\odot)<14.5$ and $0<\log(b/\textrm{kpc})<1.5$. Regions of interest were progressively refined onto a smaller grid spacing using the {\sc HyperQuadTree} algorithm kindly provided by Maarten Breddels. About 6 levels of refinement, starting from a 9x9 grid, are enough that the 7 largest values of $\sub{\super{D}{I}}{KL}$ converged to neighboring points in parameter space. The point in parameter space producing the largest $\sub{\super{D}{I}}{KL}$ was then used as $\vect{a}_0$ to calculate $\sub{\super{D}{II}}{KL}$ for all the parameter values explored in Step I. Equation \eqref{eq:kld2interp} was applied to the results of Step II to derive confidence intervals.

\section{The mock stellar halo}
\label{sec:mockhalo}
This work uses the mock Galactic stellar halo from \shh, which was generated by integrating stars from 153 progenitor satellites (Plummer spheres) as test particles in a spherical isochrone potential with total mass $M=2.7\times10^{12}\ M_\odot$ and scale radius $b=8$ kpc. The sizes and orbits of the progenitors were chosen to conform as far as possible with observations of the real MW satellites (more details in \shh). This mock halo is composed entirely of accreted satellites, some of which are fully phase-mixed (especially toward the center) and some not.

\subsection{Stellar tracers}
\begin{figure}
\includegraphics[width=0.5\textwidth]{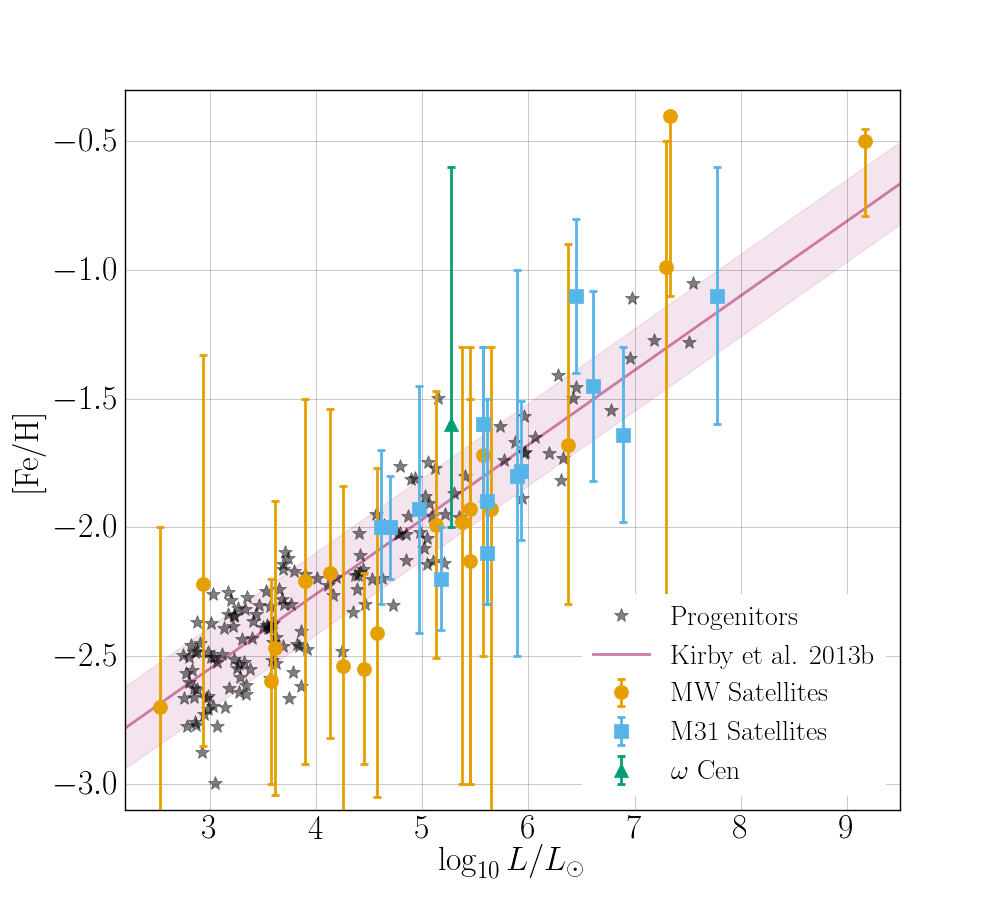}
\caption{The metallicity-luminosity relation for satellites in the Milky Way (yellow circles), Andromeda (blue squares), and Omega Centauri (green triangle), along with the relation from \citet{2013ApJ...779..102K} (Equation \ref{eq:zlum}; magenta line with $1\sigma$ spread) used to generate metallicities for the mock halo progenitors (gray stars). Vertical error bars show the approximate measured spread of metallicities per object (given in Table \ref{tbl:RRLcounts}). \label{fig:zlum}}
\end{figure}

\begin{figure}
\includegraphics[width=0.5\textwidth]{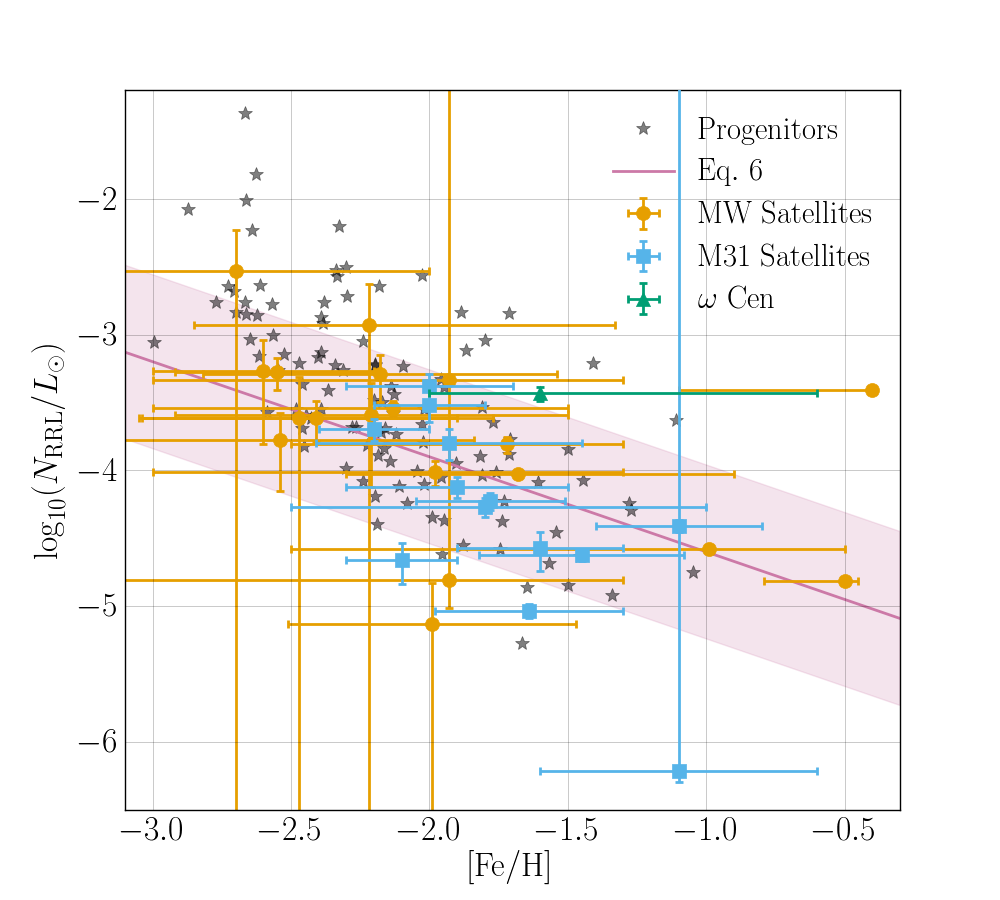}
\caption{Number of RRLe per solar luminosity as a function of mean metallicity for satellites of the Milky Way (yellow circles), Andromeda (blue squares), and Omega Centauri (green triangle). Horizontal error bars show the metallicity range for each object given in Table \ref{tbl:RRLcounts} (equivalent to the vertical error bars in Figure \ref{fig:zlum}). Vertical error bars are based on Poisson errors in the number of RRLe ($\sqrt{N_{\mathrm{RR}}}$), and extend infinitely upward in the case of lower limits (usually incomplete spatial coverage). The progenitors of the mock stellar halo (gray stars) are drawn from Equation \ref{eq:nrrl} (magenta line with $1\sigma$ spread), based on the metallicities generated with Equation \ref{eq:zlum}). \label{fig:NRRL}}
\end{figure}

To determine how many stars are in each stream, and to calculate the expected error on the astrometric parameters, some assumption must be made about the stellar populations in the building blocks forming the streams. Instead of doing a full population model, I choose a specific tracer population and estimate how many stars of that type would be expected based on the global properties of the satellite. Two types of bright tracers are considered here: KIII giants (as in \shh) and RR Lyrae (RRLe). KIII giants are very bright and relatively common, while RRLe, though relatively scarce, are bright standard candles whose distances can be determined very accurately. Both are good candidates for follow-up from the ground to complement the PMs measured by Gaia with accurate radial velocities (for faint KIII giants) or distances (for RRLe).

For spectroscopic follow-up measurements of the radial velocity, I continue to assume one KIII giant with $M_v = 1$ per 40$M_\odot$ of stellar mass.  

For RRLe distance follow-up, the number of tracers depends on the metallicity as well as the luminosity of the progenitor satellite. The number of RRLe per main-sequence star is expected to be larger for low-metallicity populations because the ratio of the time spent burning helium to the time spent on the main sequence is larger at lower metallicity \citep[i.e. low-metallicity stars of the right mass to end up as RRLe will spend a larger fraction of their lifetime in this phase than high-metallicity stars; see e.g.][]{2015ApJ...808...50M}. I use the luminosity-metallicity relation for dwarf galaxies from \citet{2013ApJ...779..102K},\begin{equation}
\label{eq:zlum}
\langle [\textrm{Fe}/\textrm{H}] \rangle = -1.68 +0.29 \log \frac{L_V}{10^6 L_\odot},
\end{equation}
to calculate a centroid of metallicity for each progenitor, and then draw the actual average metallicity for that object from a Gaussian with that centroid and spread 0.16 dex (also as measured by \citeauthor{2013ApJ...779..102K}).

For distance-measuring purposes, the quantity of interest is the number of RRLab and RRLc variable stars as a function of luminosity and metallicity, since these types have period-luminosity models of sufficient quality to determine distances to high precision (V. Scowcroft, private communication). These tracers have been observed in many satellite galaxies of both the MW and M31. Based on tabulated counts of RRLab/c from the sources listed in Table \ref{tbl:RRLcounts}, I derive a rough estimate for the number of RRLab/c expected per solar luminosity as a function of metallicity. As seen in Figure \ref{fig:NRRL} and expected based on theory, more metal-rich satellites tend to have fewer RRLe for their luminosity, and the trend is roughly a power-law. A least-squares fit without taking error bars into account gives the relationship
\begin{equation}
\label{eq:nrrl}
\log_{10} \frac{N_{RRL}}{L_\odot} = -0.70 \langle [\textrm{Fe}/\textrm{H}] \rangle - 5.3
\end{equation}
for the trend; the average spread around the trend is about 0.64 dex. Given a metallicity from Equation \eqref{eq:zlum}, the estimated number of RRLe per progenitor is drawn from a Gaussian with this spread around the value given for the trend.\footnote{Some of the satellites with metallicities and spreads from \citet{2011ApJ...727...78K} have updated values in \citet{2013ApJ...779..102K}. The result for Equation \eqref{eq:nrrl} is virtually the same and well within the spread, and the difference is too small to affect the results given here.}

This strategy for estimating the number of RRLe uses a metallicity-luminosity relation in place of an age-metallicity relation for the satellites because a satellite with a given luminosity could have a variety of star formation histories (which are not simulated in this model) and therefore a range of different ages, for which the infall time is only a lower bound. A key anchor point in determining the number of RRLe at the high metallicity end is the measurement for the LMC, which is a young object still forming stars; an older object of the same luminosity could have a lower metallicity. The relation derived here would over-estimate the metallicity, and hence under-estimate the number of RRLe, for such an object. On the other hand the largest progenitors in our mock halo are closer to the luminosity of Sagittarius, which lies above the relation in Figure \ref{fig:NRRL} by a few times the scatter. All but one of such objects in the mock halo are fairly old (i.e. have infall times larger than 8 Gyr), and all are assigned [Fe/H] between -1.5 and -1 (much lower than Sgr). One is a very recent infall so this object's metallicity should probably be higher, but the derived relation for the number of RRLe works out estimating that all these objects have about the same or a few less RRLe than Sgr. Without a more detailed model of the histories of the building blocks of the mock halo, this coarse-grained approach is the best that can be done; though the large scatter in the number of RRLe allows for some variation, there are still some individual objects with metallicities and numbers of RRLe that are inconsistent with their infall times.

The absolute visual magnitude is taken to be 0.65 for all the RRLe, based on the average value for halo RRLe measured by Hipparcos \citep{1998ApJ...492L..79T}.  The color of RRLe is metallicity-independent to good enough approximation that it is used in the literature to estimate reddening by dust along different lines of sight (which is ignored in this work). Based on the value for field RRLe quoted in \citet{1995AJ....109..588M} and references therein, the $V-I$ color is taken to be 0.58.  As described in the next section, these values for the absolute magnitude and color are used to estimate uncertainties on the proper motions and radial velocities from the Gaia and ground-based error models, while the distance uncertainty is based on estimates for how well the period-luminosity relation is normalized rather than the accuracy with which the apparent magnitude is measured. In other words, for the purpose of measuring distances, it is not assumed that all RRLe are perfect standard candles.

\begin{deluxetable*}{llllllcll}
\tablecaption{RR Lyrae and stellar metallicity ranges observed in satellites of the Milky Way and Andromeda}
\tablehead{
\colhead{} & \colhead{} & \colhead{} & \colhead{} & \colhead{} & \colhead{} &\colhead{Lower} & \colhead{} & \colhead{} \\
\colhead{Name} & \colhead{$M_V$} & \colhead{$\langle$[Fe/H]$\rangle$} & \colhead{[Fe/H]${}_\mathrm{min}$} & \colhead{[Fe/H]${}_\mathrm{max}$} & \colhead{$N_{\mathrm{RR}}$} & \colhead{Limit?}  & \colhead{Reference for $N_{\mathrm{RR}}$} & \colhead{Reference for [Fe/H] range} } 
\startdata
LMC & -18.1 & -0.5 & -0.79 & -0.45 & 22651 & - & \citet{2009AcA.59.1S} & \cite{2008AJ....135..836C} \\
Sag & -13.5 & -0.4 & -1.1 & -0.4 & 8400 & - & \cite{2001AA...375..909C} & \cite{2007ApJ...670..346C} \\
Carina & -9.1 & -1.72 & -2.5 & -1.3 & 58 & - & \cite{1995AAS..112..407K} & \cite{2006ApJ...651L.121H} \\
Draco & -8.8 & -1.93 & -3.0 & -1.3 & 131 & - & \cite{1995AAS..112..407K} & \cite{2011ApJ...727...78K} \\
Sculptor & -11.1 & -1.68 & -2.3 & -0.9 & 221 & - & \cite{1995AAS..112..407K} & \cite{2004ApJ...617L.119T} \\
UrsaMinor & -8.8 & -2.13 & -3.0 & -1.5 & 82 & - & \cite{1995AAS..112..407K} & \cite{2011ApJ...727...78K} \\
Fornax & -13.4 & -0.99 & -2.5 & -0.5 & 515 & - & \cite{2002AJ....123..840B} & \cite{2006AA...459..423B} \\
Sextans & -9.3 & -1.93 & -3.3 & -1.3 & 7 & * & \cite{2014ApJ...781...22Z} & \cite{2011MNRAS.411.1013B} \\
BootesI & -6.3     & -2.55 & -2.92 & -2.18 & 15 & - & \cite{IAU:9176910} & \cite{2010ApJ...723.1632N} \\
CVI & -8.6 & -1.98 & -3.0 & -1.3 & 23 & - & \cite{IAU:9176910} & \cite{2011ApJ...727...78K} \\
CVII & -4.9 & -2.21 & -2.92 & -1.5 & 2 & - & \cite{IAU:9176910} & \cite{2011ApJ...727...78K} \\
Coma & -4.1 & -2.6 & -3.0 & -2.2 & 2 & - & \cite{IAU:9176910} & \cite{2011ApJ...727...78K} \\
LeoIV & -5.8 & -2.54 & -3.24 & -1.84 & 3 & - & \cite{IAU:9176910} & \cite{2011ApJ...727...78K} \\
UMII & -4.2 & -2.47 & -3.04 & -1.9 & 1 & - & \cite{IAU:9176910} & \cite{2011ApJ...727...78K} \\
UMI & -5.5 & -2.18 & -2.82 & -1.54 & 7 & - & \cite{IAU:9176910} & \cite{2011ApJ...727...78K} \\
Hercules & -6.6 & -2.41 & -3.05 & -1.77 & 9 & - & \cite{IAU:9176910} & \cite{2011ApJ...727...78K} \\
LeoT & -8.0 & -1.99 & -2.51 & -1.47 & 1 & - & \cite{IAU:9176910} & \cite{2011ApJ...727...78K} \\
Segue1 & -1.5 & -2.7 & -3.4 & -2.0 & 1 & - & \cite{2011ApJ...733...46S} & \cite{2010ApJ...723.1632N} \\
Segue2 & -2.5 & -2.22 & -2.85 & -1.33 & 1 & - & \cite{2013AJ....146...94B} & \cite{2013ApJ...770...16K} \\
AndI & -11.7 & -1.45 & -1.82 & -1.08 & 98 & - & \cite{IAU:9176910} & \cite{2010ApJ...711..671K} \\
AndII & -12.4 & -1.64 & -1.98 & -1.3 & 72 & - & \cite{IAU:9176910} & \cite{2010ApJ...711..671K} \\
AndIII & -10.0 & -1.78 & -2.05 & -1.51 & 51 & - & \cite{IAU:9176910} & \cite{2010ApJ...711..671K} \\
AndV & -9.1 & -1.6 & -1.9 & -1.3 & 10 & - & \cite{IAU:9176910} & \cite{2011MNRAS.417.1170C} \\
AndVI & -11.3 & -1.1 & -1.4 & -0.8 & 111 & - & \cite{IAU:9176910} & \cite{2011MNRAS.417.1170C} \\
AndIX & -8.1 & -2.2 & -2.4 & -2.0 & 30 & - & \cite{IAU:9176910} & \cite{2010MNRAS.407.2411C} \\
AndX & -7.6 & -1.93 & -2.41 & -1.45 & 15 & - & \cite{IAU:9176910} & \cite{2010ApJ...711..671K} \\
AndXI & -6.9 & -2.0 & -2.2 & -1.8 & 15 & - & \cite{IAU:9176910} & \cite{2010MNRAS.407.2411C} \\
AndXIII & -6.7 & -2.0 & -2.3 & -1.7 & 17 & - & \cite{IAU:9176910} & \cite{2010MNRAS.407.2411C} \\
AndXVI & -9.2 & -2.1 & -2.3 & -1.9 & 9 & - & \cite{IAU:9176910} & \cite{2009MNRAS.400.1472L} \\
AndXIX & -9.2 & -1.9 & -2.3 & -1.5 & 31 & - & \cite{IAU:9176910} & \cite{2008ApJ...688.1009M} \\
AndXXI & -9.9 & -1.8 & -2.5 & -1.0 & 42 & - & \cite{IAU:9176910} & \cite{2009ApJ...705..758M} \\
NGC147 & -14.6 & -1.1 & -1.6 & -0.6 & 36 & * & \cite{2010ApJ...708..293Y} & \cite{2010ApJ...711..361G} \\
OmegaCen & -8.35 & -1.6 & -2.0 & -0.6 & 69 & - & \cite{2007AJ....133.1447W} & \cite{2007ApJ...654..915S}\\
\enddata
\tablecomments{$M_V$: absolute visual magnitude; $\langle$[Fe/H]$\rangle$: mean stellar metallicity; $N_{\mathrm{RR}}$: number of observed RR Lyrae. [Fe/H]${}_{\mathrm{min}, \mathrm{max}}$ indicate the range of observed stellar metallicities: symmetric ranges denote instances where a metallicity dispersion was reported; the span is 2$\sigma$ in these cases. Asymmetric ranges were estimated by eye from histograms. An asterisk (*) in the ``Lower Limit" column indicates that the observations from which RRLe were identified only partially covered the target galaxy.}
\label{tbl:RRLcounts}
\end{deluxetable*}

%

\begin{figure}
\includegraphics[width=0.5\textwidth]{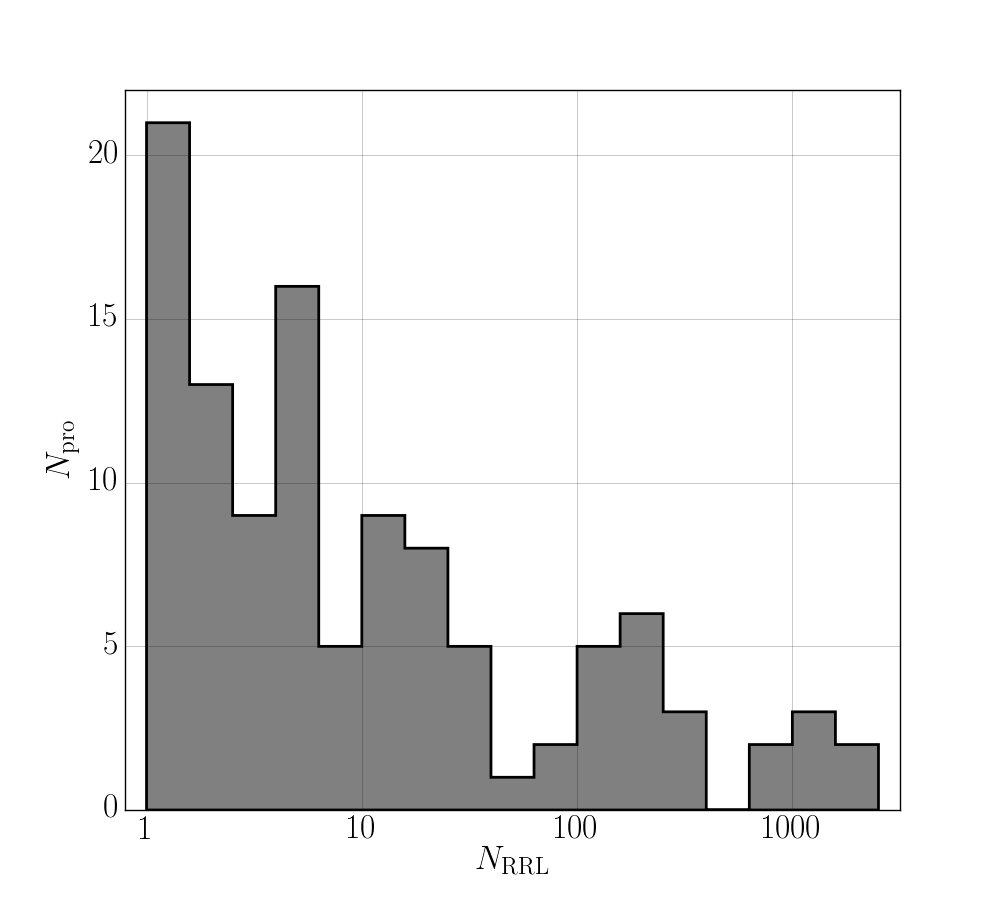}
\caption{Number of RRLe in each progenitor in the mock stellar halo. Of the 153 progenitors, 110 have at least 1 RRL and 45 have more than 10. \label{fig:RRL_hist}}
\end{figure}

\subsection{Error modeling}
\begin{figure}
 \plotone{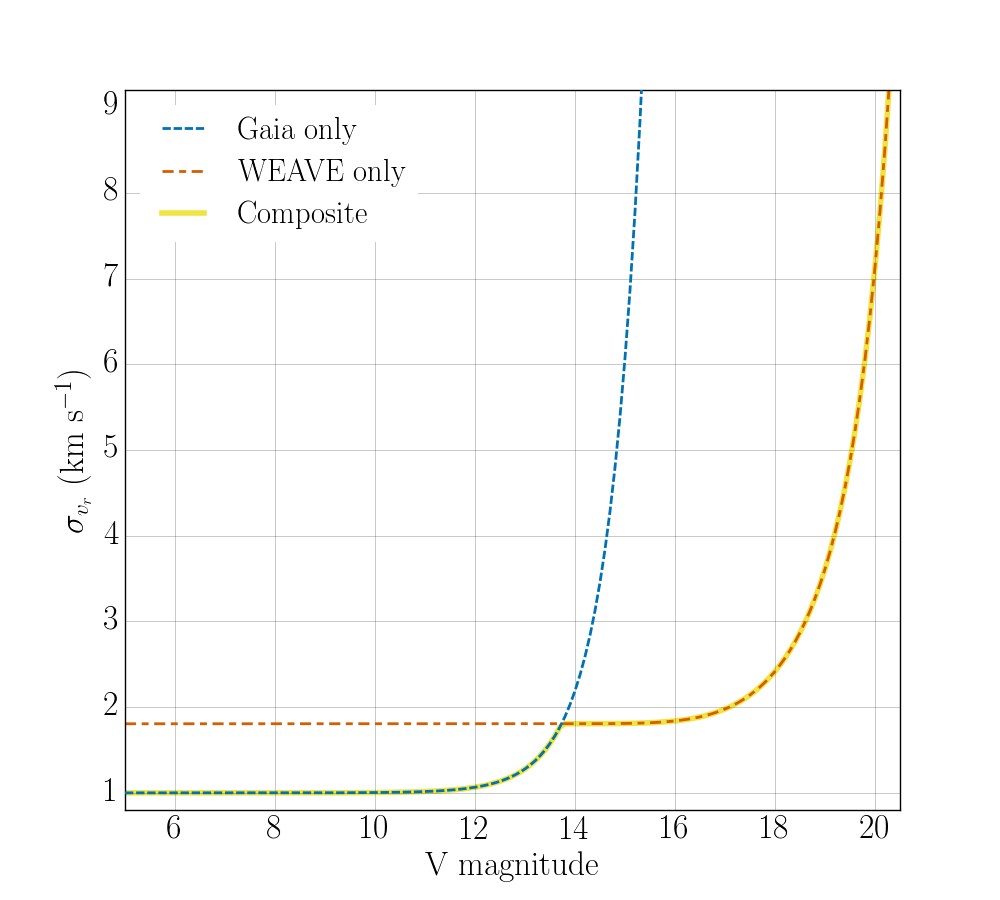}
\caption{Fitting functions for determining the magnitude of the radial-velocity error, $\sub{\sigma}{RV}$, based on the Gaia post-launch performance (blue dashed curve) and projected WEAVE performance as of 2014 (red dot-dashed curve) as a function of the apparent $V$ magnitude. The yellow curve highlights the composite used in this work, which selects the smaller of the two error estimates for a given $V$. Gaia's RVs are more accurate for bright stars ($V\lesssim 13.5$).}
\label{fig:ffuncs}
\end{figure}

The error-convolution procedure is similar to that described in Section 3.4 of \shh, but now using a post-launch Gaia error model that accounts for the different number of observations as a function of ecliptic latitude and longitude (courtesy C. Mateu and the Gaia Challenge Workshop).  Error convolution is applied to both halos, the KIII giants and the RRLe, selecting only stars for which Gaia will measure PMs. 

For the KIII giants, we assume as before that Gaia will determine parallaxes to at least 20 percent, consistent with estimates for end-of-mission photometric distances, but we now allow for ground-based radial velocity measurements for faint stars by calculating the projected error from both Gaia and ground-based follow-up surveys, and choosing the smaller of the two. For the projected ground-based RV error we use an error-magnitude fitting formula based on WEAVE forecasts kindly provided by S. Trager; this curve was accurate as of 2014 but recent advances have improved the forecasted errors so that the floor is closer to 1 km \unit{s}{-1}. The projected errors for 4MOST (less than 2 km \unit{s}{-1} to 20th magnitude; see \url{https://www.4most.eu/cms/science/}) are also better at faint magnitudes than the curve adopted here. Given the assumed RV error, the result is that for stars brighter than about $V=13.5$ the Gaia RV error is used, while the ground-based RV error is used for $13.5<V<20$. This transition point is about 1 magnitude brighter than projected by pre-launch performance estimates for Gaia's RVS. We also implicitly assume that all KIII halo stars in the Gaia catalog are followed up from the ground, consistent with what is currently planned for 4MOST (A. Helmi, priv. comm.) but not necessarily for WEAVE. Figure \ref{fig:ffuncs} shows the two fitting formulae and the composite curve that results from this definition. 

For the RRLe, all stars in the sample have \emph{distances} (not parallaxes) measured to 2\%, again optimistically assuming that all the RRLe in the Gaia proper-motion catalog will be followed up. Because of their intrinsic variability, spectroscopic radial velocities for these stars are expected to be somewhat less accurate than nonvariable stars; the final accuracy is in the range 5-10 km \unit{s}{-1} (B. Sesar, priv. comm.). The results here conservatively assume an RV error of 10 km \unit{s}{-1} based on the assumption that RRLe not observed by Gaia to this accuracy will be followed up from the ground; improving the RV accuracy to 5 km \unit{s}{-1} did not affect the results.

In all cases, the error-convolved observables are obtained for each star by drawing a value from a Gaussian distribution centered at the true value with a width the size of the error. 

\subsection{Selecting stars for fitting}
\label{subsec:selections}

After convolving with the errors stars were selected from the halo for data quality. For the RRLe, which are quite bright but fairly scarce, no selections were made other than to require measurements for all 6 phase-space coordinates.  For the more numerous KIII giants, three different strategies for error-selection were adopted and are described below. The statistics for the selected stars---number of stars, number of progenitors, and average distance---are summarized in Table \ref{tbl:stats}. For comparison, the 4MOST survey, which will cover a bit less than half the sky, expects to follow up between 1 and 3 million halo giants.

The first strategy was to keep the same maximum transverse velocity error as in \shh, $\sub{\sigma}{v_t}<18.15$ km \unit{s}{-1}, and make no cut on the RV error. This sample adds all the stars that satisfy the PM and parallax error requirements but were too faint ($V>17.3$) to have an RV measurement from Gaia. 

The second strategy was to require that the absolute transverse velocity error for all stars be smaller than the largest RV error in the sample, which is how the maximum acceptable transverse velocity error was determined in \shh. Due to the better quality ground-based RVs this limit decreases to $\sub{\sigma}{v_t}<7.1$ km \unit{s}{-1}. Again, there is no selection on the RV error. 

The final strategy was to require that \emph{each} star's transverse velocity error be roughly equal to or smaller than its RV error, specifically $\sigma_{v_t} \leq 1.3 \sigma_{v_r}$. This is the strictest criterion of the three for the sample with ground-based RVs, but is much less strict for the Gaia-only sample where the RV errors are much larger.

The last step in selecting the stars for fitting is to downsample the deeper regions of the potential to prevent overcrowding in action space from producing a spurious maximum corresponding to the Kepler potential, using the strategy described in Section 4 of \shh. In short, a trial energy is calculated for each star using a logarithmic potential with circular velocity (normalization) $v_c = \langle |\vect{L}| /r \rangle$, and stars with trial energies less than a cutoff value are discarded. The values for $v_c$, the constant potential offset, and the cutoff energy are the same as in \shh. After applying this energy selection to a sample of KIII giants with maximum transverse velocity error $X$, the resulting sample is called {\sc Gaia\_plus\_Xkms}. The sample with the per-star variable cutoff is {\sc Gaia\_plus\_strict}. The error and energy selections are performed using the error-convolved positions and velocities, but parallel samples are constructed that contain the same stars with unconvolved positions and velocities to isolate the effect of the observational errors. The tags {\sc \_er} and {\sc \_ne} appended to the sample names indicate whether the positions and velocities for the stars in a sample are error-convolved or not, respectively.

\begin{table*}
\begin{center}
\caption{Mock halo selections}
 \begin{tabular}{llllllll}
& & $\langle d \rangle$, & $\langle d_\odot \rangle$, & $\langle d \rangle$, & $\langle d_\odot \rangle$, & & $\sub{N}{pro},$ \\
 Sample & $N_*$ & no ec & no ec & with ec & with ec & $\sub{N}{pro}$ & $n_* > 100$ \\
\hline
\hline

Full KIII giant mock halo & 6 765 774 & \ph{1}5.74 & \ph{1}9.70 & N/A & N/A & 153 & 106 \\

\hline

$\sub{\sigma}{v_t}<18$ km \unit{s}{-1}, Gaia+Ground & 3 684 518 & \ph{1}5.66 & \ph{1}9.30 & \ph{1}5.16 & \ph{1}8.22 & 152 & \ph{1}89 \\

$\sub{\sigma}{v_t}<18$ km \unit{s}{-1}, Gaia Only & 3 621 989 & \ph{1}5.40 & \ph{1}9.04 & \ph{1}4.94 & \ph{1}8.00 & 149 & \ph{1}84 \\

{\sc Gaia\_plus\_18kms} & \ph{0 }569 169 & \ph{1}8.18 & \ph{1}9.87 & \ph{1}7.59 & \ph{1}9.07 & 127 & \ph{1}52 \\
{\sc Gaia\_only\_18kms} & \ph{0 }518 034 & \ph{1}6.61 & \ph{1}8.16 & \ph{1}6.30 & \ph{1}7.62 & 124 & \ph{1}45 \\

\hline

$\sub{\sigma}{v_t}<7$ km \unit{s}{-1}, Gaia+Ground & 2 514 261 & \ph{1}6.23 & \ph{1}9.76 & \ph{1}5.59 & \ph{1}8.54 & 152 & \ph{1}79 \\
$\sub{\sigma}{v_t}<7$ km \unit{s}{-1}, Gaia Only & 2 451 172 & \ph{1}5.86 & \ph{1}9.39 & \ph{1}5.28 & \ph{1}8.21 & 147 & \ph{1}72 \\
{\sc Gaia\_plus\_7kms} & \ph{0 }380 629 & \ph{1}9.57 & 10.6 & \ph{1}8.78 & \ph{1}9.57 & 124 & \ph{1}47 \\
{\sc Gaia\_only\_7kms} & \ph{0 }326 894 & \ph{1}7.36 & \ph{1}8.00 & \ph{1}6.97 & \ph{1}7.37 & 114 & \ph{1}38 \\

\hline

$\sub{\sigma}{v_t}<1.3 \sigma_{v_r}$, Gaia+Ground&\ph{0 }604 897 & \ph{1}7.50 &10.5  & \ph{1}6.67 & \ph{1}9.12 & 133 & \ph{1}47 \\
$\sub{\sigma}{v_t}<1.3 \sigma_{v_r}$, Gaia Only& 3 288 360 & \ph{1}5.49 & \ph{1}9.40 & 4.98 & 8.26 & 149 & \ph{1}78 \\
{\sc Gaia\_plus\_strict} & \ph{0 0}91 497 & 14.1 & 14.9 & 12.5 & 13.1 & \ph{1}96 & \ph{1}29 \\
{\sc Gaia\_only\_strict} & \ph{0 }433 024 & \ph{1}6.71 & \ph{1}8.87 & \ph{1}6.33 & \ph{1}8.20 & 119 & \ph{1}41 \\

\hline
\hline

RR Lyra mock halo & \ph{0 0}12 935 & \ph{1}6.252 & 10.190 & \ph{1}6.255 & 10.193 & 110 & \ph{1}21\\

\end{tabular}
\tablecomments{Size and distance statistics for stars in the mock halo after the various error cuts described in Section \ref{sec:mockhalo}. Listed are the total number of stars in each sample $N_*$, mean galactocentric distance $\langle d \rangle$ in kpc, mean heliocentric distance$\langle d_\odot \rangle$ in kpc, number of progenitors \sub{N}{pro}, and number of progenitors with more than 100 stars. The distances are calculated two ways: using the positions of the selected stars without errors (``no ec'') and using the error-convolved positions (``with ec''). {\sc Gaia\_only\_18kms} are the samples used in \shh\ reprocessed with the updated Gaia error model. }
\label{tbl:stats}
\end{center}
\end{table*}


\subsection{Characteristics of the fitting samples}
\label{subsec:samples}

\begin{figure}
\includegraphics[width=0.45\textwidth]{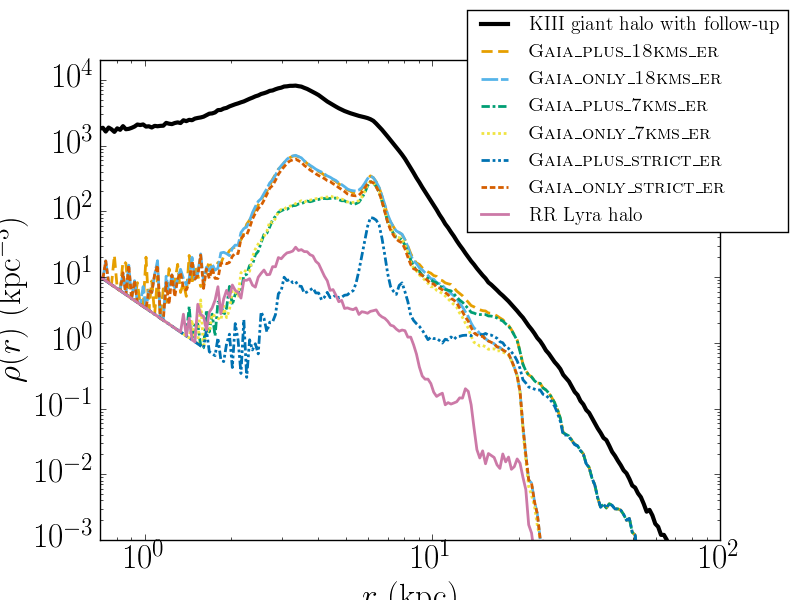}
\caption{Galactocentric radial density distribution of stars in the various mock halo selections used for fitting, constructed from the error-convolved parallaxes. Information about the different fitting samples is summarized in Table \ref{tbl:stats}.}
\label{fig:fitdist}
\end{figure}

\begin{figure*}
\begin{tabular}{ccc}
\includegraphics[width=0.32\textwidth]{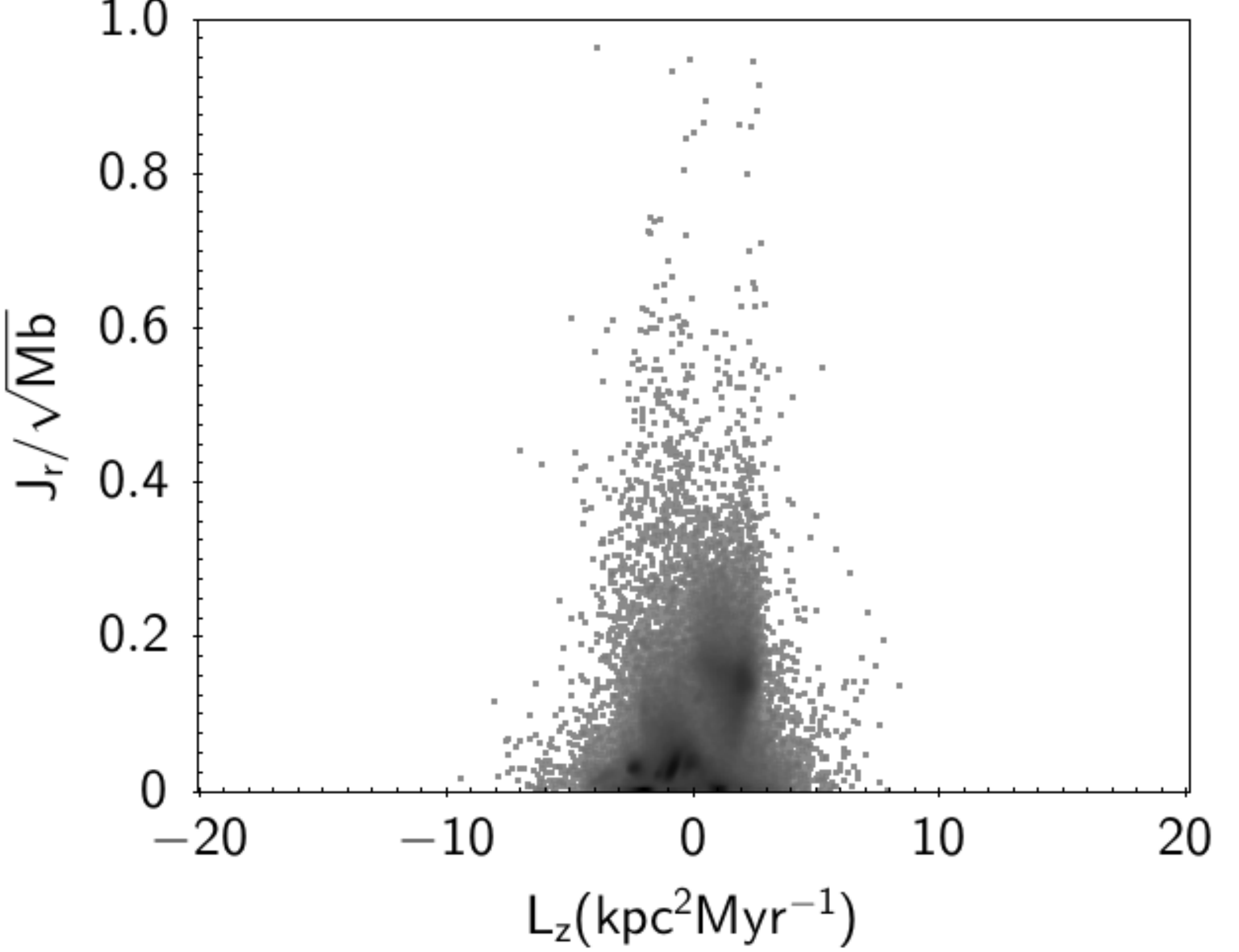} & \includegraphics[width=0.32\textwidth]{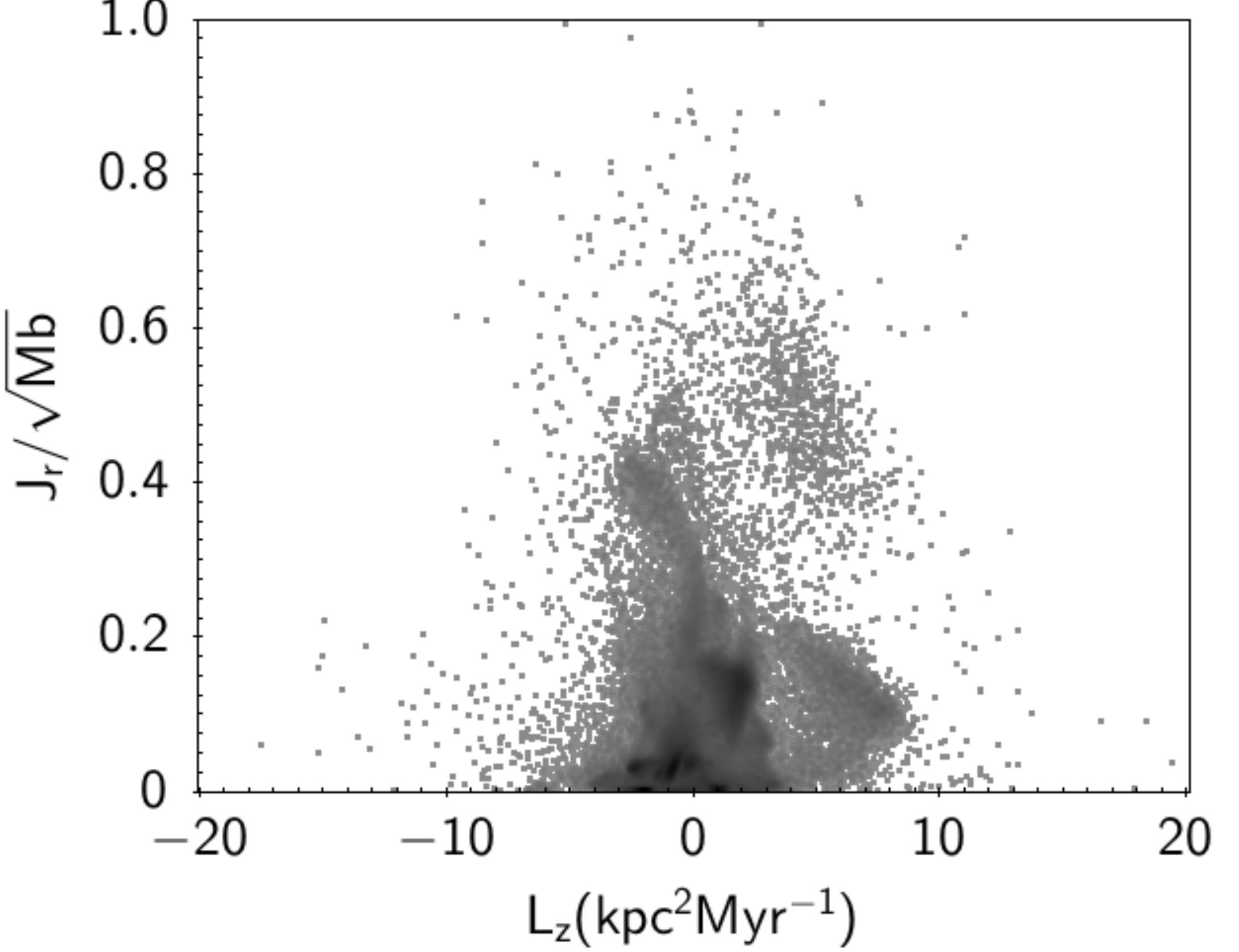} & \includegraphics[width=0.32\textwidth]{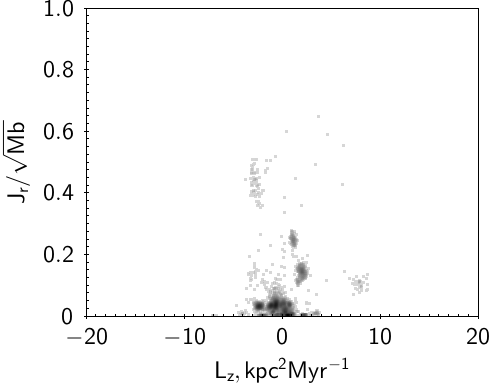}
\end{tabular}
\caption{Action-space distributions for {\sc Gaia\_only\_7kms\_er} (left), {\sc Gaia\_plus\_7kms\_er}, (center), and the entire RRLe halo including observational errors (right).}
\label{fig:ActionsRRLand7kms}
\end{figure*}

The density distributions of the fitting samples, as a function of Galactocentric distance, are shown in Figure \ref{fig:fitdist}. The effect of selecting stars by estimated energy generally chooses more distant stars; interestingly, selecting stars from the followed-up samples by requiring star-by-star consistency in the velocity errors also tends to pick out more stars at large distances. Rather counterintuitively, as the error requirements on the transverse velocity become stricter, the average distance of a sample increases. This effect is mainly due to the steep density profile of the mock halo: there are many stars at small Galactocentric distances but large heliocentric distances (i.e. on the other side of the bulge from the Sun) that will therefore have fairly large PM errors and be discarded.

Figure \ref{fig:fitdist} also illustrates the impact of the follow-up RVs to extend the distance range of the fitting samples, from about 20 kpc maximum to about 50 kpc. Unlike the followed-up sample, the RRLe sample is not distance-limited by lack of observations for some stars, but instead by our assumptions about the assembly history of the mock halo: we preferentially place the largest satellites on orbits with small apocenters to mimic dynamical friction, so many of the satellites orbiting far out in the halo have at most only a few RRLe. There is also a straightforward downsampling effect: the number of RRLe per satellite is far lower than the number of KIII giants and the density profile of the sampled apocenters is very steep (leading to $\rho \propto r^{-3.5}$) so above roughly 25-30 kpc there are far fewer stars in general.

The three KIII giant samples with follow-up primarily differ by how many stars at small radii are included in the sample. The samples using only KIII giants observed completely with Gaia, except for the {\sc strict} sample, mainly follow the followed-up sample until the edge of Gaia's capability to measure RVs. The effect of the {\sc strict} error selection is quite different for the followed-up sample than for the Gaia-only one, so that there is very little difference between {\sc Gaia\_only\_18kms} and {\sc Gaia\_only\_strict}. The RRLe sample, which includes everything in the RRLe mock halo for which the error models predict all 6 phase-space coordinates will be measured, has a markedly different distribution at smaller radii thanks to the inverse dependence of RRLe occurrence with metallicity and hence stellar mass. The RRLe sample thus has the advantage that it naturally downsamples the larger satellites in the interior of the halo, whereas this was done manually (and without membership information) using an approximate energy cut for the KIII giants.

Figure \ref{fig:ActionsRRLand7kms} shows the projected distribution of the true actions (calculated using the input potential for the mock halo) for one of the error selections made to the KIII giant halo, compared to the RRLe halo. Comparing the left column (KIII halo without follow-up) to the center one (KIII halo with follow-up) shows that follow-up does add some extra streams, and therefore clumps, to each sample. Comparing with the final panel (the RRLe halo) makes it clear that the primary culprit that is blurring the error distributions is the distance error: all the stars in the left and center panels have RV errors better than 7 km \unit{s}{-1}, while the RRLe halo assumes RV errors of 10 km \unit{s}{-1}. Thus while the RV errors for the RRLe are somewhat larger than for the KIII giants, the RRLe distance errors are much smaller, leading to a dramatically clearer picture of the action distribution.

\section{Results}
\label{sec:results}

\begin{figure}
\begin{center}
\plotone{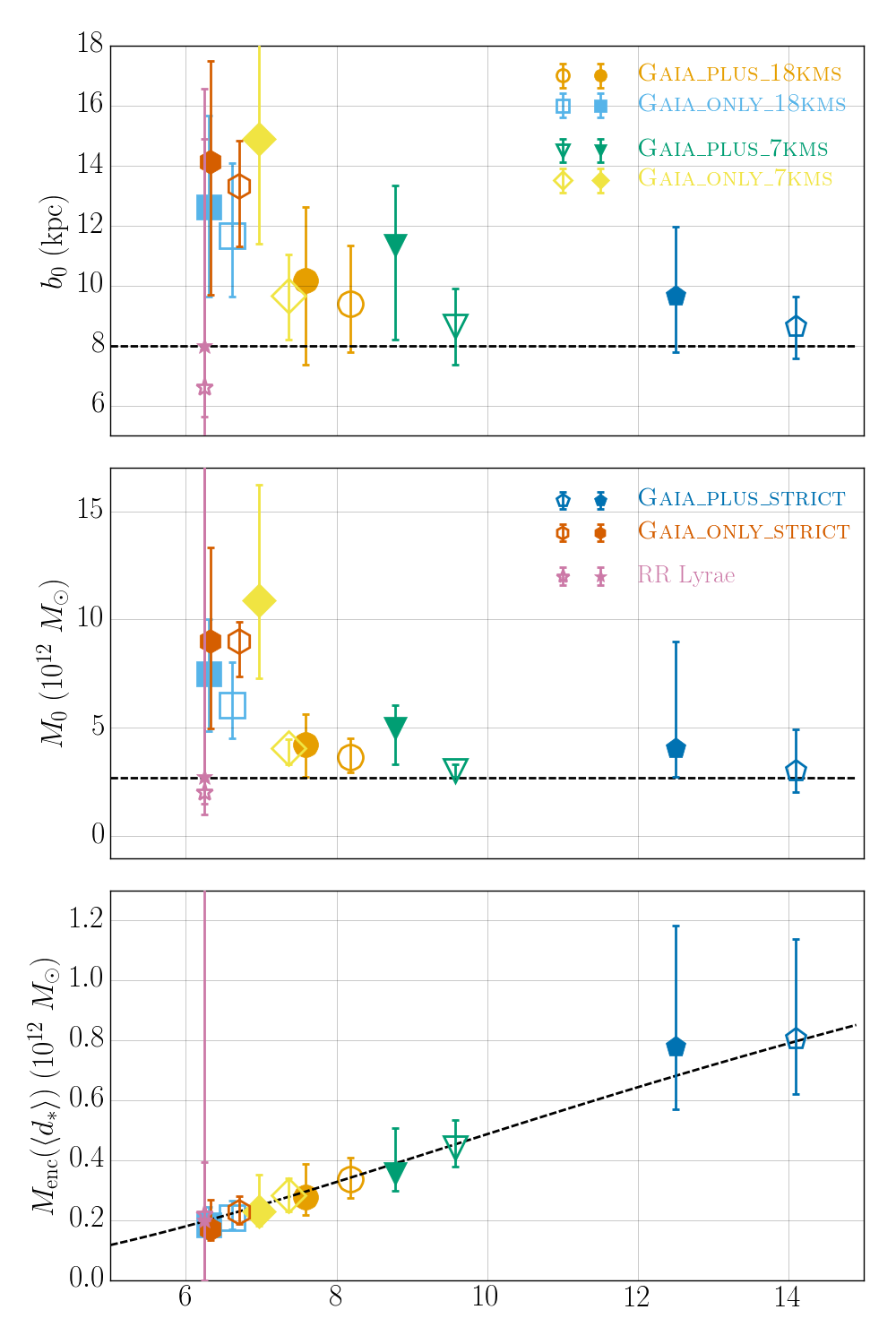}
\caption{Values of the recovered potential parameters, and their uncertainties, for the various samples as a function of their average Galactocentric distance, $\langle d_* \rangle$. Results from samples without error convolution are marked with open symbols; results from samples with error convolution are shown as closed symbols. Marker sizes are proportional to the log of the number of stars in the sample. The top panel shows how recovery of the scale radius $b$ improves with increasing average distance; the middle panel shows recovery of the total mass $M$. The bottom panel shows the recovered enclosed mass at the average stellar distance, $\sub{M}{enc}(\langle d_* \rangle )$, which is well-constrained for all samples. }
\label{fig:resultsVsDist}
\end{center}
\end{figure}

Figure \ref{fig:resultsVsDist} shows the recovered parameters from all the fits with different samples as a function of their average Galactocentric distance, $\langle d_* \rangle$. The values plotted in this figure are included in Tables \ref{tbl:results} and \ref{tbl:resultsNE}. The results using error-convolved data are shown as filled symbols; those from ``perfect" data (the same set of points prior to error convolution) are shown with open symbols. Pairs sharing a shape and color were selected in the same way. The error bars are the extrema of the $\super{\sub{D}{KL}}{II}=1/2$ contour in each parameter, analogous to $1\sigma$ uncertainties as discussed in Section \ref{sec:kld}. The average distance for the error-convolved and non-convolved samples of KIII giants differs thanks to parallax error. The error-convolved sample always has a smaller average distance as a result of Lutz-Kelker bias \citep{1973PASP...85..573L}, an effect of measuring parallax (inverse distance) instead of distance. During error convolution of the mock halo the parallaxes are drawn from symmetric Gaussian distributions, but the corresponding distribution in distance space is quite asymmetric, and is not peaked at the distance corresponding to the true parallax but at a smaller value, with larger deviations for larger parallax errors. The stacked distribution of ``observed" (i.e. error-convolved) distances for a given selection, which samples the composite distribution of parallaxes, thus peaks at a distance lower than the true mean. The RRLe distances measured using the period-luminosity relation are not subject to this bias.

Samples with larger average distances tend to do better at recovering the nominal parameters used to describe the potential: the scale radius $b$ (top panel) and the total mass $M$ (center panel). The bottom panel shows that in fact all samples recover accurately the enclosed mass at $\langle d_* \rangle$,
\begin{equation}
\label{eq:enclosedMass}
\sub{M}{enc}(\langle d_* \rangle) \equiv \frac{M \langle d_* \rangle^3}{\sqrt{b^2 + \langle d_* \rangle^2}\left(b^2 + \sqrt{b^2 + \langle d_* \rangle^2}\right)^2},
\end{equation}
as discussed in \shh.  Thus samples for which this enclosed mass is a larger fraction of the total mass (i.e. those with larger average distance) also do better at recovering $b$ and $M$, which are individually degenerate. The size of the uncertainty on the recovered $\sub{M}{enc}$ roughly corresponds to the number of stars in the fitting sample (which is indicated by the size of the markers), while the uncertainties on $b$ and $M$ also reflect the degree of degeneracy for that particular fitting sample.

For the KIII giant halo, observational errors (primarily parallax error) cause both $M$ and $b$ to be over-estimated ({\sc Gaia\_only\_18kms}, blue squares, and {\sc Gaia\_only\_strict}, red hexagons, in Figure \ref{fig:resultsVsDist}). Observational error biases the results by blurring the action-space clumps, and parallax errors in particular tend to stretch the clumps in radial action ($J_r$) more than in angular momentum ($L$ and $L_z$). In this toy potential only $J_r$ changes with the potential parameters, and there is some degeneracy between the stretching in $J_r$ caused by observational errors and the way the clumps deform as the parameters change, that in this case leads to the slight over-estimation of $M$ and $b$. This degeneracy is worse if the stars are all at small Galactocentric distance since this limits the range of possible $J_r$ for the clumps; with a larger distance (and therefore $J_r$) range the degree of stretching from manipulating the potential parameters differs more from one stream (clump) to another and improves the fit. 

However observational error does not explain all the bias in the samples with the most stars at small distances, which still over-estimate both parameters. Even without observational errors, the streams still have a finite size in action space that is larger for the more massive streams, which have the most weight in the fit since each star is considered independently. The remaining bias comes from the attempt to adjust the potential parameters to reduce the extent of these few largest clumps in $J_r$ below their intrinsic size. The stretching of the clumps with $M$ and $b$ is $J_r$ dependent, so the degeneracy is worst when all the largest clumps have similar mean actions, again arguing for choosing a fitting sample that has the widest possible range of orbits.

Being slightly more selective on velocity errors makes little difference (compare {\sc Gaia\_only\_7kms}, yellow diamonds), but adding follow-up observations from the ground (compare {\sc Gaia\_plus\_18kms}, orange circles, and {\sc Gaia\_plus\_7kms}, green triangles) improves performance significantly. The followed-up samples do as well at recovering $M$ and $b$ when including observational errors (filled symbols) as the corresponding samples without follow-up can do with perfect data (open symbols). Error selections that significantly reduce the sample size (e.g. {\sc Gaia\_plus\_strict}, blue pentagons) decrease the bias at the cost of increasing the uncertainties.

The results from fitting the RRLe (in Figure \ref{fig:resultsVsDist}, pink stars) are an interesting counterpoint to the KIII giant halo in a few ways. First, for this sample the observational errors are so small and well-behaved (thanks mostly to the small distance errors) that the results are nearly identical whether observational errors are included or not. Second, although the stars in this sample have an average Galactocentric distance of only 6.25 kpc, the high quality of the data and the more even distribution in radius lead to a very accurate estimate of the potential parameters. However, thanks to the very small number of stars in this data set, which is less than a tenth the size of the KIII giant samples, the uncertainties on the best fit values are very large. 

In these tests the power of the RRLe distances is muted somewhat by the relatively steep profile from which the satellites' orbits are drawn, so that relatively few satellites orbit at large distances. Those that do are too small to contain more than a handful of RRLe, because satellites the size of Sagittarius or Fornax are also preferentially placed on orbits with small apocenters to mimic dynamical friction. This may be an overly conservative set of assumptions, since several present-day Milky Way satellites at distances larger than 25 kpc contain tens to hundreds of RRLe. If the halo's building blocks included similar satellites on distant orbits, the constraining power of RRLe would increase correspondingly as they would then trace streams on a wider variety of orbits. This would lead to an action-space in which the clumps of stars were better separated from each other, whereas in this mock halo many of the clusters overlap and can be confused with one another when comparing different potentials, especially for clumps that are not well resolved. For a distribution with less overlap, there would be less confusion, reducing the size of the uncertainties on the parameters even if the total sample size remained the same.

\section{Discussion}
\label{sec:discussion}

The results presented here point toward the dual importance of gathering 6D data on stars in \emph{distant} streams (for instance by following up faint stars in Gaia), even if the distances are only marginally well known, and of getting \emph{high-quality distances} (e.g. from RRLe in streams) even if this is only possible for stars in relatively nearby streams. Both types of observational effort will significantly improve our ability to measure the Galactic mass profile, at least for the method tested here.  It is clear from recent work \citep[e.g.][Sanderson, Hartke, \& Helmi in prep]{2013MNRAS.433.1826S,2013MNRAS.436.2386L,2014ApJ...795...94B} that single tidal streams have limited ability to place constraints on the Galactic potential no matter how well their stars' positions and velocities are measured, and that the best strategy is to target streams with a wide variety of orbits. Therefore observing numerous tracers like KIII giants at large distances complements the precise distances available for less numerous tracers like RRLe, since the KIII giants' wider reach brings a wider variety of possible orbits into the sample. Combining the two types of tracers will allow a more precise picture of the potential nearby while giving a more complete picture of its mass and shape overall.

It is worth noting that the fitting method used to obtain these results does not require membership information, i.e. assigning stars to a particular stream, and does not construct a generative model of the action-space distribution, but merely maximizes the amount of information in the distribution in a statistical sense. This allows rapid fitting of a large sample of stars, but the tradeoff is that there is no way to properly account for observational errors, including Lutz-Kelker bias, in the fit itself. A method that uses a true likelihood function, i.e. constructs a generative model for the distribution of stars in each stream and compares it to the observed stars, has the potential to treat the observational errors properly, but those available \citep[e.g.][]{2013MNRAS.433.1826S,2014MNRAS.443..423S,2014ApJ...795...95B,2015ApJ...803...80K} require assigning stars to a particular stream. The complementarity of methods for determining the MW potential that do and do not require membership information in the context of the Gaia dataset is the subject of a forthcoming paper inspired by the Gaia Challenge workshop series (\url{http://tinyurl/gaiachallenge}), in which a total of eight different methods will be compared (Sanderson et al., in prep). 

The primary drawback to augmenting the Gaia catalog with RRLe distances, from the perspective of the fitting method I discuss here, is the difficulty of obtaining light-curves for large numbers of halo RRLe, which are sparsely distributed over the sky. Space-based IR observations like those with Spitzer are so far vastly preferable to ground-based follow-up thanks to Spitzer's well-calibrated, uniform, precise photometry, but such observations are expensive in terms of observing time: multiple observations are required to place each RRL on the $P$-$L$ relation, and the average density of halo RRLe on the sky is less than 1 per degree, so each star requires its own set of repeat pointings. The SMHASH survey \citep{2013sptz.prop10015J}, taking advantage of the relatively high density of RRLe in the known Sagittarius and Orphan streams, will still observe less than two hundred RRLe. Ideally this would be a prime task for future ground-based surveys, but although some recent work indicates that the same 2\% accuracy might be achieved from the ground \citep{2014arXiv1404.4870K}, this is still controversial; the calibration issues associated with ground-based observations have not yet been fully solved. 

Another intriguing possibility for RRLe in the more distant future is the prospect of combining distances from the Large Synoptic Survey Telescope (LSST)'s high-cadence observations of RRLe lightcurves \citep{2012AJ....144....9O,2015ApJ...812...18V} with proper motions measured by the WFIRST High-Latitude Survey \citep{2013arXiv1305.5422S} which will span 2000 square degrees at high Galactic latitudes in six IR bands, with sensitivity and uniformity comparable to Hubble over a field of view 100 times larger \citep{2013arXiv1305.5425S}. Under the current plan, WFIRST would revisit the high-latitude fields (located in the southern hemisphere to intersect LSST) at least 5 times over the course of the survey with a cadence that would allow measurement of proper motions to $\sim100\mu$as precision for stars as faint as 27th magnitude in the $J$ band. If current efforts to better connect pulsation models to optical and near IR lightcurves \citep[for example][]{2015ApJ...808...50M} lead LSST to deliver 2\% RRLe distances from multiband photometry, we would then someday be able to push our knowledge of the six-dimensional phase space distribution of the Milky Way's halo beyond the faint limit of Gaia, out to its very edge. 

These results also suggest that prospects for using RRLe to determine the halo's mass profile out to large distances will be significantly affected by the number and metallicity of the progenitors of distant tidal streams in the Galaxy. This particular mock halo has no massive streams at large distances, which limits the effectiveness of the RRLe as fit tracers. In cosmological simulations the assembly histories of MW-mass galaxies vary widely, leading to great diversity in the chemodynamics of their stellar halos \citep{2010MNRAS.406..744C}, and we do not know yet fully know what degree of structure our MW's stellar halo contains. However, combining multiple tracers to improve fit performance should help in the case where few RRLe are found in distant streams. Similarly, the chemical compositions of the progenitor galaxies themselves also differ from each other, though the metallicity spreads overlap one another, adding extra dimensions of information that could also be used to constrain the Galactic potential. I intend to explore both these strategies in future work.

\begin{table*}
 \begin{center}
\caption{Fit results for samples with errors\label{tbl:results}}
\renewcommand{\arraystretch}{1.5}
\begin{tabular}{lrccccccc}
 
Sample & $N_*$ & $\langle d_* \rangle$ & $\super{\sub{M}{true}}{enc}(\langle d_* \rangle)$ &$\super{M_0}{enc}$ & $M_0$ ($\sub{M}{true}=2.7$) & $b_0$ ($\sub{b}{true}=8.0$)& $\sub{\super{D}{I}}{KL}(\vect{a}_0)$ & $\super{\sub{D}{KL}}{I}(\sub{\vect{a}}{true})$ \\

\hline
\hline
{\sc Gaia\_plus\_18kms\_er}  & 569 169 & 7.59 &  0.294 & $ 0.277\substack{+ 0.110\\-0.058}$ & $  4.22\substack{+  1.41\\ -1.48}$ & $ 10.18\substack{+  2.45\\ -2.82}$ &  1.96 &  1.91\\
{\sc Gaia\_only\_18kms\_er}  & 518 034 & 6.30 &  0.199 & $ 0.186\substack{+ 0.059\\-0.046}$ & $  7.50\substack{+  2.50\\ -2.63}$ & $ 12.63\substack{+  3.04\\ -2.99}$ &  1.99 &  1.90 \\

\hline

{\sc Gaia\_plus\_7kms\_er} & 380 629 & 8.78 &  0.388 & $ 0.355\substack{+ 0.152\\-0.056}$ & $  4.96\substack{+  1.08\\ -1.62}$ & $ 11.34\substack{+  1.99\\ -3.14}$ &  2.01 &  1.97\\
{\sc Gaia\_only\_7kms\_er} & 326 894 & 6.97 &  0.247 & $ 0.231\substack{+ 0.120\\-0.048}$ & $ 10.90\substack{+  5.30\\ -3.60}$ & $ 14.90\substack{+  3.58\\ -3.51}$ &  2.09 &  1.96\\

\hline

{\sc Gaia\_plus\_strict\_er} & \ph{3}91 497 & 12.5 &  0.679 & $ 0.778\substack{+ 0.404\\-0.208}$ & $  4.07\substack{+  4.91\\ -1.33}$ & $\ph{1}9.65\substack{+  2.32\\ -1.87}$ &  1.65 &  1.63\\
{\sc Gaia\_only\_strict\_er} & 433 024 & 6.33 &  0.202 & $ 0.169\substack{+ 0.099\\-0.036}$ & $  8.98\substack{+  4.36\\ -4.02}$ & $ 14.10\substack{+  3.40\\ -4.40}$ &  1.92 &  1.78\\

\hline
\hline

RR Lyrae & \ph{3}12 935 & 6.25 & 0.196 & $0.201\substack{+2.18\\-0.20}$ & $2.74\substack{+62.2\\-1.23}$ & $\ph{1}7.99\substack{+6.89\\-2.36}$ & 1.09 & 1.02\\

\end{tabular}
\end{center}

\tablecomments{$\super{\sub{M}{true}}{enc}(\langle d_* \rangle)$: Enclosed mass at mean radius in true potential. $\super{M_0}{enc}$: Enclosed mass at mean radius in best-fit potential. $M_0, b_0$: best-fit total mass and scale radius. $\sub{\super{D}{I}}{KL}(\vect{a}_0)$: Maximum value of the KLD found in Step I. $\super{\sub{D}{KL}}{I}(\sub{\vect{a}}{true})$: KLD of Step I for the true values of the potential parameters.  All masses are in units of $10^{12} M_\odot$; all distances are in kpc. Error bars indicate the extrema of the two-dimensional 68\% confidence contours.}
\end{table*}

\begin{table*}
 \begin{center}
\caption{Fit results for samples without errors\label{tbl:resultsNE}}
\renewcommand{\arraystretch}{1.5}
\begin{tabular}{lrccccccc}
 
Sample & $N_*$ & $\langle d_* \rangle$ & $\super{\sub{M}{true}}{enc}(\langle d_* \rangle)$ &$\super{M_0}{enc}$ & $M_0$ ($\sub{M}{true}=2.7$) & $b_0$ ($\sub{b}{true}=8.0$)& $\sub{\super{D}{I}}{KL}(\vect{a}_0)$ & $\super{\sub{D}{KL}}{I}(\sub{\vect{a}}{true})$ \\

\hline
\hline

{\sc Gaia\_plus\_18kms\_ne} & 569169 &  8.18 &  0.340 & $ 0.336\substack{+ 0.073\\-0.063}$ & $  3.65\substack{+  0.88\\ -0.71}$ & $9.39\substack{+  1.95\\ -1.62}$ &  2.44 &  2.35 \\

{\sc Gaia\_only\_18kms\_ne} & 518034 &  6.61 &  0.221 & $ 0.208\substack{+ 0.059\\-0.037}$ & $  6.04\substack{+  2.01\\ -1.51}$ & $11.65\substack{+  2.42\\ -2.01}$ &  2.55 &  2.38 \\
   
\hline

{\sc Gaia\_plus\_7kms\_ne} & 380629 &  9.57 &  0.451 & $ 0.441\substack{+ 0.094\\-0.063}$ & $  3.02\substack{+  0.31\\ -0.29}$ & $8.66\substack{+  1.25\\ -1.29}$ &  2.46 &  2.44 \\

{\sc Gaia\_only\_7kms\_ne} & 326894 &  7.36 &  0.277 & $ 0.282\substack{+ 0.059\\-0.053}$ & $  4.07\substack{+  0.42\\ -0.73}$ & $9.65\substack{+  1.39\\ -1.44}$ &  2.62 &  2.55 \\

\hline

{\sc Gaia\_plus\_strict\_ne} & 91497 &  14.10 &  0.794 & $ 0.806\substack{+ 0.332\\-0.184}$ & $  3.02\substack{+  1.93\\ -0.99}$ & $8.66\substack{+  0.99\\ -1.09}$ &  2.17 &  2.15 \\

{\sc Gaia\_only\_strict\_ne} & 433024 &  6.71 &  0.228 & $ 0.227\substack{+ 0.053\\-0.038}$ & $  8.98\substack{+  0.93\\ -1.61}$ & $13.30\substack{+  1.52\\ -1.99}$ &  2.53 & 2.36  \\

\hline
\hline

RR Lyrae & \ph{3}12 935 & 6.25 & 0.196 & $0.221\substack{+2.87\\-0.173}$ & $2.04\substack{+62.9\\-1.02}$ & $6.61\substack{+9.94\\-2.08}$ & 1.06 & 1.03  \\

\end{tabular}
\end{center}

\tablecomments{$\super{\sub{M}{true}}{enc}(\langle d_* \rangle)$: Enclosed mass at mean radius in true potential. $\super{M_0}{enc}$: Enclosed mass at mean radius in best-fit potential. $M_0, b_0$: best-fit total mass and scale radius. $\sub{\super{D}{I}}{KL}(\vect{a}_0)$: Maximum value of the KLD found in Step I. $\super{\sub{D}{KL}}{I}(\sub{\vect{a}}{true})$: KLD of Step I for the true values of the potential parameters.  All masses are in units of $10^{12} M_\odot$; all distances are in kpc. Error bars indicate the extrema of the two-dimensional 68\% confidence contours.}
\end{table*}

\acknowledgements
The author is supported by an NSF Astronomy and Astrophysics Postdoctoral Fellowship under award AST-1400989. She is indebted to 
Maarten Breddels (Groningen) for use of the grid resampling code;
Andreas K\"upper (Columbia), Cecilia Mateu (UNAM), and the Gaia Challenge for contributions to the error-convolution code;
Kathryn Johnston (Columbia), Vicky Scowcroft (Carnegie) and Branimir Sesar (MPIA) for explaining the current state-of-the-art in RRLe observations;
Scott Trager (Groningen) for discussions about WEAVE;
David Spergel (Princeton) for consultation on astrometry with WFIRST;
{\u Z}eljko Ivezi{\' c} (U. Washington) for information about LSST observations of RRLe;
the Kapteyn Computer Group (Groningen) for support of the computing cluster used to carry out this work;
and
Amina Helmi (Groningen) for information about 4MOST and general support, and the Stream Team NYC (Columbia, NYU, and Yale) for general discussions that shaped this paper. The colorblind-friendly color scheme used to create the figures is from Okabe \& Ito, 2002 (\url{http://jfly.iam.u-tokyo.ac.jp/color/}).

\bibliography{gbfu}

\end{document}